\documentclass[11pt]{article} \textwidth 170mm \textheight 225mm
\usepackage{amsfonts}
\usepackage{amssymb}
\usepackage{dcolumn}
\usepackage{amsthm}
\usepackage{amsmath}
\usepackage{multicol}
\usepackage{newlfont}
\usepackage{newlfont}
\usepackage{multicol}
\usepackage{bm}
\usepackage{txfonts}
\usepackage{graphics}
\usepackage{graphicx}
\usepackage{color}
\usepackage{indentfirst}
\begin{document}

\title{First-principles calculations on temperature-dependent elastic constants of rare-earth intermetallic  compounds:\\ YAg and YCu \footnote{The work is supported by the National Natural Science Foundation
of China (11074313) and and Project No.CDJXS11102211 supported by the Fundamental Research Funds for the Central Universities of China. }}
\author{Rui Wang\footnote{Tel: +8613527528737; E-mail: rcwang@cqu.edu.cn.}, Shaofeng Wang,  Xiaozhi Wu, and Yin Yao\\
{\small  { Institute for Structure and Function and
department of physics, Chongqing University, }}\\ {\small {
Chongqing 400044, People's Republic of China. }} }

\date{}

\maketitle
\begin{abstract}
\noindent We present the temperature-dependent elastic constants of two ductile rare-earth intermetallic compounds YAg and YCu with CsCl-type B2 structure by using a first-principles approach. The elastic moduli as a function of temperature are predicted from the combination of static volume-dependent elastic constants obtained by the first-principles total-energy method with density functional theory and the thermal expansion obtained by the first-principles phonon calculations with density-functional perturbation theory. The comparison between our calculated results and the available experimental data for Ag and Cu provides good agreements. In the calculated temperature $0-1000K$, the elastic constants of YAg and YCu follow a normal behavior with temperature that those decrease with increasing temperature, and satisfy the stability conditions for B2 structures. The Cauchy pressure for YAg and YCu as a function of temperature is also discussed, and our results mean that YAg and YCu  become more ductile while increasing temperature.
\end{abstract}

\vskip 0.1in{\small   } \vskip 0.2in

\vskip 0.1in

 \vskip 0.3in

\baselineskip 20pt


\section{Introduction}

The rare-earth intermetallic compounds typically possess high strength and stiffness, low specific weight
corrosion resistance, and hot strength superior to ordinary metals \cite{Lazar,Gumbsch}. However, their room
temperature brittleness and poor fracture toughness severely restrict their applications. Recently, a large family of rare-earth intermetallics has been discovered to be ductile and
tough at room temperature \cite{Gschneidner}, in some cases exceeding 20\% ductility in tension. These alloys have the B2 structure in space
group Pm3m (number 221) and chemical formula
RM, where R denotes a rare-earth element, and M denotes a late transition metal or an
early p-element.  Many theoretical and experimental works \cite{Morris, Chen1,Shi,Wu1,Wu2,Wang2010, Chen2,Zhang,Zhang1,Russell} have been done to
understand the mechanism of high ductility at room temperature, such as elasticity, defect properties, phase stability, dislocations, stacking fault, electronic structure and density of states (DOS), etc. However, the properties and mechanism have not been understood completely hitherto.

To get a better understanding of the anomalous ductility of RM intermetallics, more fundamental investigations of temperature-dependent properties such as thermal expansion, heat capacity, and thermoelasticity, etc., are obviously required. In general, elastic properties of a solid are very important because they are closely associated with various fundamental solid-state properties such as interatomic potentials, equation of state, and phonon spectra. The temperature dependence of the elastic constants of a material is important for predicting and understanding the mechanical strength, stability, and phase transitions of a material \cite{Gulsern}. However, there are few literatures focused on the temperature-dependent moduli of RM intermetallics. The first-principles methods based on density functional theory (DFT) can compute the single-crystal elastic moduli at zero-temperature accurately, but treating the corresponding temperature dependence of elastic constants is still a formidable challenge \cite{Orlikowski}. More recently, Wang et. al. \cite{Wang,Shang} proposed a simple quasistatic approach to determine the elastic constants at finite temperatures by using first-principles calculations, and the excellent agreement between the calculated results and experimental data is found. Their approach is based on the fact that the temperature dependence of elastic constants mainly results from volume expansion with increasing temperature \cite{Gulsern,Ledbetter,Swenson}.

In the present work, we apply the first-principles quasistatic approach to investigate the temperature-dependent elastic constants for the novel intermetallics YAg and YCu. The static volume-dependent elastic constants are obtained by using the DFT total-energy calculations combined with the method of homogeneous deformation.  A first-principles quasiharmonic approach \cite{Togo2010} is applied to predict the phonon free energy and volume versus temperature by using  density-functional perturbation theory (DFPT). To benchmark
the reliability results of the presented method, we have also calculated the temperature-dependent elastic constants for Ag and Cu, and the comparison between our predicted results and the available experimental data provides good agreements. In addition, the  ductility of YAg and YCu as a function of temperature has been investigated by using the temperature-dependent elastic modulus.

\section{Theory}

In this section, we present the necessary theory for the calculation of the temperature dependence of cubic crystal elastic moduli in B2 structures. For programming the calculation of the elastic constants, we consider a procedure of homogeneous deformation of a bulk-crystal. With the displacement given as $u_{i}= x_{i}-a_{i}$ $(i=1, 2, 3)$, between the initial configuration $a_{i}$ and the strained configuration $x_{i}$, the deformation applied to the crystal is described as the deformation gradient matrix $J_{ij}=\partial x_{i}/ \partial a_{j}$. Then, we may define the Lagrangian strain $\eta_{ij}=\frac{1}{2}(J_{ik}J_{kj}-\delta_{ij})$ \cite{Thurston}. Thus, the isothermal elastic constants $C_{ijkl}^{T}$ are defined as the derivative with respect to the Lagrangian strain on  the Helmholtz free energy Eq.(\ref{Ftotal}) at constant temperature \cite{Brugger},
\begin{equation}\label{Cijkl}
C_{ijkl}^{T}=\frac{1}{V}\frac{{\partial}^{2}F}{\partial \eta_{ij}
\partial \eta_{kl}}\bigg|^{T}_{\eta '},
\end{equation}
where $V=V(T)$ is the equilibrium volume at temperature $T$, and $\eta '$ indicates that all other stains are held fixed. This is the general way to compute the finite temperature elastic constants in principle. However, this procedure is quite cumbersome at the present since it involves calculation of the second derivatives of $F_{\mathrm{vib}}$.
For cubic materials, there are there independent elastic constants $C_{11}^{T}$, $C_{12}^{T}$, and $C_{44}^{T}$ (in Voigt notation) which describe the elastic behavior completely. A more convenient set for computations are $C_{44}^{T}$ and two linear combinations $B^{T}$ and $\mu^{T}$. The bulk modulus
\begin{equation}\label{Bulk}
B^{T}=(C_{11}^{T}+2C_{12}^{T})/3,
\end{equation}
is the resistance to deformation by a uniform hydrostatic pressure; the shear modulus
\begin{equation}\label{shear}
\mu^{T}=(C_{11}^{T}-C_{12}^{T})/2,
\end{equation}
is the resistance to shear deformation across the (110) plane in the $\langle110\rangle$ direction, and $C_{44}$ is the resistance to shear deformation across the (100) plane in the  $\langle010\rangle$ direction. The bulk modulus $B$ is determined from the Vinet equation of state \cite{Vinet}. The shear modulus $\mu$ is calculated from volume-conserving orthorhombic strain
\begin{equation}\label{strain1}
\eta(\delta)=\left( \begin{array}{ccc}
                      \delta & 0 & 0 \\
                      0 & \delta & 0 \\
                      0 & 0 & (1+\delta)^{-2}-1
                    \end{array}
\right),
\end{equation}
and the Helmholtz free energy related to this strain is
\begin{equation}\label{Fs1}
F(V,\delta)=F(V,0)+6\mu^{T} V \delta^{2}+O(\delta^{3}),
\end{equation}
where $F(V,0)$ is the free energy of the unstrained structure.
We use a volume-conserving tetragonal strain to determine $C_{44}^{T}$,
\begin{equation}\label{strain2}
\eta(\delta)=\left( \begin{array}{ccc}
                      0 & \delta & 0 \\
                      \delta & 0 & 0 \\
                      0 & 0 & \delta^{-2}/(1-\delta^{-2})
                    \end{array}
\right),
\end{equation}
which leads to the free energy change
\begin{equation}\label{Fs2}
F(V,\delta)=F(V,0)+2C_{44}^{T} V \delta^{2}+O(\delta^{4}),
\end{equation}

In the usual way and following Moriarty et al \cite{Moriarty}, the Helmholtz free energy $F$ for a crystal at temperature $T$ and  volume $V$
can be separated as
\begin{equation}\label{Ftotal}
F(V, T)=E_{\mathrm{0}}(V)+F_{\mathrm{vib}}(V, T)+F_{\mathrm{el}}(V,T),
\end{equation}
where $E_{\mathrm{0}}(V)$ is the zero-temperature total energy of the electronic ground state, $F_{\mathrm{vib}}(V, T)$ is the vibrational free energy which  comes from the phonon contribution, and where $F_{\mathrm{el}}$ represents the thermal electronic contribution to free energy from finite temperature. Here $V$ is taken as the volume per unit cell and $F$ is the free energy per unit cell. The total hydrostatic pressure in the crystal can be calculated by solving
\begin{equation}
P(V,T)=-\frac{\partial F(V,T)}{\partial V}\Bigg|_{T},
\end{equation}
with corresponding relations for its three components. Other thermodynamic functions can be obtained  from $F(V,T)$ similarly, e.g., entropy $S=-(\partial F/\partial T)_{V}$, internal $E=F+TS$, and the Gibbs free energy $G=F+PV$.

In usual way, the vibrational free energy $F_{\mathrm{vib}}(V, T)$ includes both quasiharmonic and anharmonic components. The anharmonic effects which represent the phonons interaction and electron-phonon coupling are so weak and can be neglected \cite{Orlikowski}. In this work, we only consider the remaining quasiharmonic phonon free energy which can be written as
\begin{equation}\label{Fph}
F_{\mathrm{vib}}(V, T)=\sum_{\mathbf{q}\lambda}\Bigg[\frac{1}{2}{\hbar\omega_{\mathbf{q}\lambda}}+{k_{B}T}\ln\bigg(1-e^{-{\hbar\omega_{\mathbf{q}\lambda}}/{k_{B}T}}\bigg)\Bigg],
\end{equation}
where $k_{\mathrm{B}}$ represents the Boltzmann constant, $\hbar$ is the reduced Planck constant, and $\omega_{\mathbf{q}\lambda}$ represents the frequency of the $\lambda$th phonon branch at wave vector $\mathbf{q}$.

With finite temperature DFT calculations \cite{Mermin}, the thermal electronic
contribution to free energy $F_{\mathrm{el}}$ is obtained from the energy and entropy contribution and given by
\begin{equation}
F_{\mathrm{el}}=E_{\mathrm{el}}-TS_{\mathrm{el}},
\end{equation}
with the bare electronic entropy $S_{\mathrm{el}}$
\begin{equation}\label{ele-entropy}
S_{\mathrm{el}}(V,T)=-k_{\mathrm{B}}\int_{0}^{\infty}n(\varepsilon,V)[f(\varepsilon)\ln f(\varepsilon) +(1-f(\varepsilon))\ln (1-f(\varepsilon))] d\varepsilon,
\end{equation}
and the thermal electronic energy  $E_{\mathrm{el}}$
\begin{equation}\label{ele-Energy}
E_{\mathrm{el}}(V,T)=\int_{0}^{\infty}n(\varepsilon,V)f(\varepsilon)\varepsilon d\varepsilon-\int_{0}^{\varepsilon_{F}}n(\varepsilon,V)\varepsilon d\varepsilon,
\end{equation}
where $n(\varepsilon,V)$, $f(\varepsilon)$, and $\varepsilon_{F}$ represent the electronic density of state (DOS), the Fermi-Dirac distribution, and the Fermi energy, respectively.

In present work, we employ a first-principles quasistatic approach, which was developed by Wang et al.\cite{Wang,Shang}, to calculate the temperature-dependent elastic constants $C_{ij}(T)$. Following the quasistatic approximation, the change of elastic properties at elevated temperatures is mainly caused by volume change due to thermal expansion, and the contributions of $F_{\mathrm{vib}}$ and $F_{\mathrm{el}}$ to these second derivations can be neglected.  In fact, the contributions due to the kinetic energy and the fluctuation of microscopic stress tensor are small and can be ignored reasonably \cite{Lutsko}.  The thermal expansion and the equilibrium volume $V$ at $T$ are obtained by using the quasiharmonic approximation in our calculation. A similar procedure was employed by K\'{a}das et al \cite{Kadas}, though they calculated thermal expansion by using Debye-type model. We obtain the temperature dependence of isothermal elastic constants $C_{ij}^{T}(T)$ by the application of the following three-step procedure. The first step is calculating the thermal expansion and the equilibrium volume  $V(T)$ at $T$ by using the first-principles quasiharmonic approach. In this procedure, one can compute volume-dependence of phonon density of states at $0K$ in a set of volume points, and the predicted equilibrium volume  $V(T)$ at $T\neq0K$ (or inversely $T(V)$ relation) is determined from fitting to the Vinet equation of state \cite{Vinet} by minimizing free energy $F$ with respect to $V$, while isothermal bulk modulus $B^{T}$ as a function of temperature $T$ is also obtained. In the second step, we obtain the  volume-dependent elastic constants $C_{ij}^{T}(V)$ at $T=0K$
 as the second derivatives of the Helmholtz free energies with respect to strain tensor by using the energy-strain relation based on Eqs.(\ref{Bulk}-\ref{Fs2}). In the third step, the calculated elastic constants from the second step at the volume $V(T)$ are approximated as those at finite temperatures, i.e., $C_{ij}^{T}(T)=C_{ij}^{T}(T(V))$. There is ample experimental evidence \cite{Anderson,Swenson, Wasserman} to support this approximation in which the temperature dependence of elastic constants are solely caused by thermal expansion.

In order to compare with experimental data, the isothermal elastic moduli $C_{ij}^{T}$ must also be transformed to the adiabatic elastic moduli $C_{ij}^{S}$ by the following relation \cite{Davies}:
\begin{equation}\label{TtoS}
C_{ij}^{S}=C_{ij}^{T}+\frac{TV}{C_{V}}\lambda_{i}\lambda_{j}
\end{equation}
where $C_{V}$ is the specific heat at constant volume and
\begin{equation}
\lambda_{i} =\sum_{k}\alpha_{k}C_{ik}^{T},
\end{equation}
with the linear thermal expansion tensor $\alpha_{k}$ . For cubic crystals, Eq. (\ref{TtoS}) simplifies to
\begin{eqnarray}
&C_{44}^{S}=C_{44}^{T}, \\
&C_{11}^{S}=C_{11}^{T}+\Delta, \\
&C_{12}^{S}=C_{12}^{T}+\Delta,
\end{eqnarray}
where \cite{Orlikowski}
\begin{equation}
\Delta=\frac{TV}{C_{V}}\alpha^{2}{B^{T}}^{2},
\end{equation}
with the volume thermal expansion coefficient $\alpha=V^{-1}{\partial V}/{T}|_{p}$. It is worth noted that the correction $\Delta$  increases with temperature.

\section{Computational details}

In present work, the static energy and the thermal electronic contribution to the Helmhotz free energy were computed by using the first-principles calculations in the framework of the density-functional theory (DFT). We employed the plane-wave basis projector augmented wave (PAW) method \cite{Blochl, Kresse4} within the generalized gradient approximation (GGA) in the  Perdew-Burke-Ernzerhof (PBE) \cite{Perdew1,Perdew2}
exchange-correlation functional as implemented in the VASP code \cite{Kresse1, Kresse2,Kresse3}.  The radial cutoffs of the PAW potentials of Ag, Cu, and Y were 1.50, 1.31 and 1.81 {\AA}, respectively. The 4$d$ and 5$s$ electrons for Ag, the 3$d$ and 4$s$ electrons for Cu, and the 4$s$, 4$p$, 4$d$ and 5$s$ electrons for Y  were treated as valence and the remaining electrons were kept frozen. The Brillouin zones (BZ) of the unit cells are represented by Monkhorst-Pack special $k$-point scheme \cite{Monkhorst}. Since high accuracy is needed to evaluate the elastic constants, the convergence of strain energies with respect to the Brillouin zone integration was carefully checked by repeating the calculations for $21\times21\times21$ and $25\times25\times25$ grid meshes at equilibrium volume at 0K, and we found at most 0.5GPa difference both for $C_{44}$ and $\mu$. Hence, we used $21\times21\times21$ in the full Brillouin zone giving 726 irreducible k-points. In addition, we used a high plane-wave energy cutoff of 600eV which is sufficient to calculate the elastic moduli accurately. The thermal electronic  energies and entropies are evaluated by using one-dimensional integrations from the self-consistent DFT calculations of electronic DOS using FD smearing as shown in Eqs. (\ref{ele-entropy}) and (\ref{ele-Energy}). Calculations of the static strain-energy were performed by the tetrahedron method with Bl\"{o}chl corrections \cite{Blochl1}, which gives a good account for total energy.

The vibrational free energy were obtained from the first-principles phonon calculations by using PHONOPY \cite{Togo2010,Togo2008,Togop} which can support VASP interface to calculate force constants directly in the framework of  density-functional perturbation theory (DFPT) \cite{Kresse5}. Phonon calculations were performed by the supercell approach. Since the chosen supercell size strongly influences on the thermal properties, we compare the vibrational free energies of $3\times3\times3$ supercell with those of $5\times5\times5$ supercell at 300K and 1000K, and find that the energy fluctuations between $3\times3\times3$ and $5\times5\times5$ supercells are less than 0.01\%. Hence, we chose the $3\times3\times3$ supercell with 54 atoms to calculate phonon dispersions. We carried out DFPT calculations on this 54 atoms supercell using PBE-GGA exchange-correlations effects and $7\times7\times7$ k-point grid meshes for BZ integrations. In order to deal with the possible convergence problems for metals, a smearing technique is employed by using the Methfessel-Paxton scheme \cite{Methfessel}, with a smearing with of 0.05eV.

We have calculated Helmholtz free energy [the right-hand side of Eq.(\ref{Ftotal})] at temperature points with a step of 1K from 0 to 1000K  at 15 volume points. At each temperature point, the equilibrium volume $V$ and isothermal bulk moduli $B^{T}$ are obtained by minimizing free-energy with respect to $V$ from fitting the integral form of the Vinet equation of state (EOS) \cite{Vinet}. Then, the volume thermal expansion coefficient was obtained by numerical differentiation for $\partial V / \partial T$.

\section{Results and discussion}

Table \ref{table} gives  our calculated lattice constants and isentropic elastic constants $C_{ij}^{S}$ at $T=0K$ and $T=300K$ at ambient pressure ($P=0\mathrm{GPa}$) in comparison with the results from the previous calculations \cite{} ($T=0K$) and experimental data at ambient conditions \cite{Morris}($P=0\mathrm{GPa}$, $T=300K$). In all cases, the comparison is quite agreeable. At $T=0K$ our calculated results show a very good agreement with the other theoretical data. For YCu, the present results at $300K$ are very closed to the experimental values obtained by Morris et al \cite{Morris} at ambient conditions. The largest discrepancy between theory and experiment is the value for $C_{44}^{S}$, where the difference is approximately 10\% for YAg.

\begin{figure}
\scalebox{0.6}[0.6]{\includegraphics{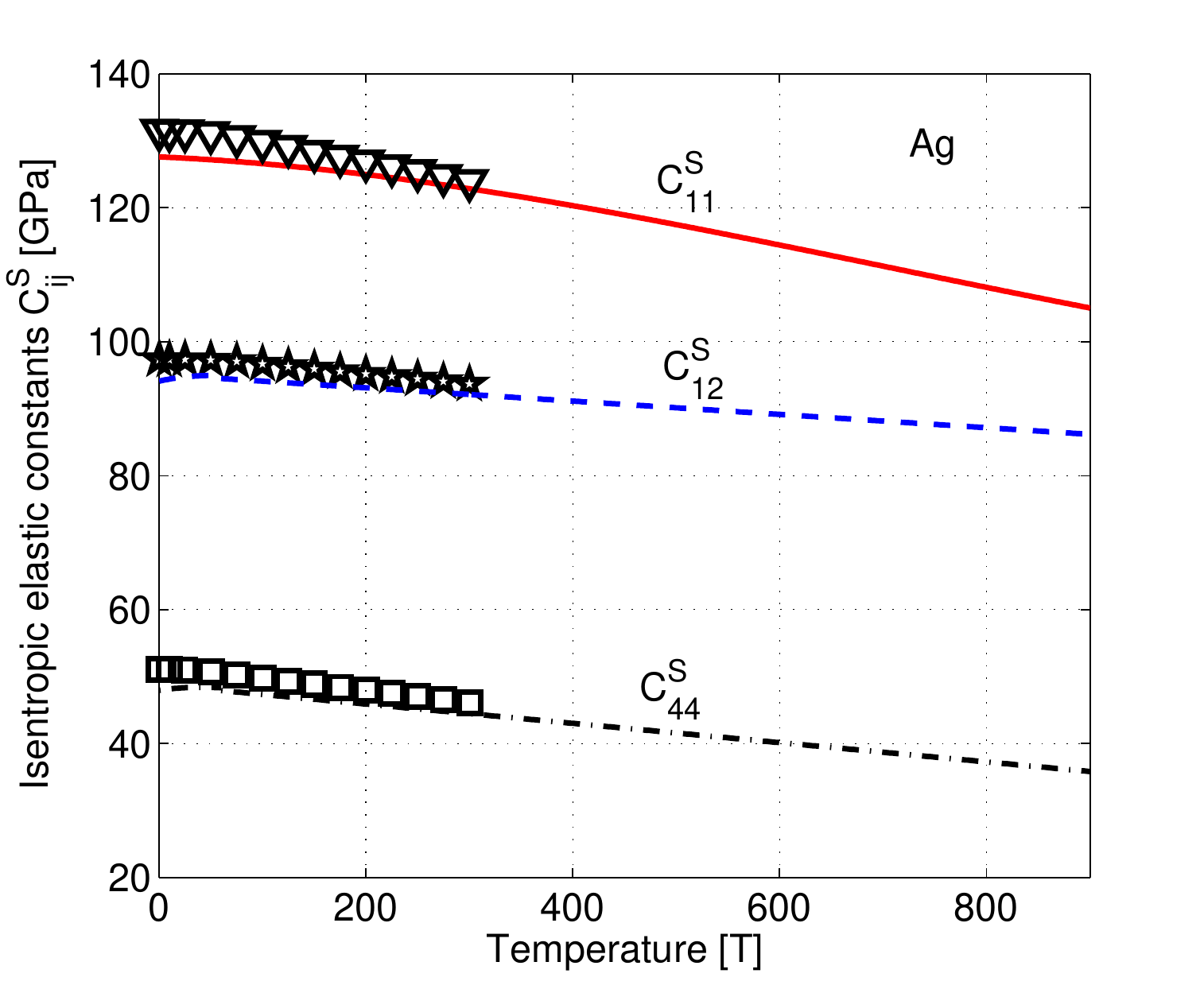}}
\caption{For single-crystal Ag, the present calculated isentropic elastic constants and corresponding experimental data. The solid, dashed, and dashed-dotted curves denote the present values of $C_{11}^{S}$, $C_{12}^{S}$ and $C_{44}^{S}$, respectively. The open symbols denote the corresponding values of ultrasonic measurements by Neighbours and Alers  \cite{Neighbours}. }
\label{Ag}
\end{figure}

\begin{figure}
\scalebox{0.6}[0.6]{\includegraphics{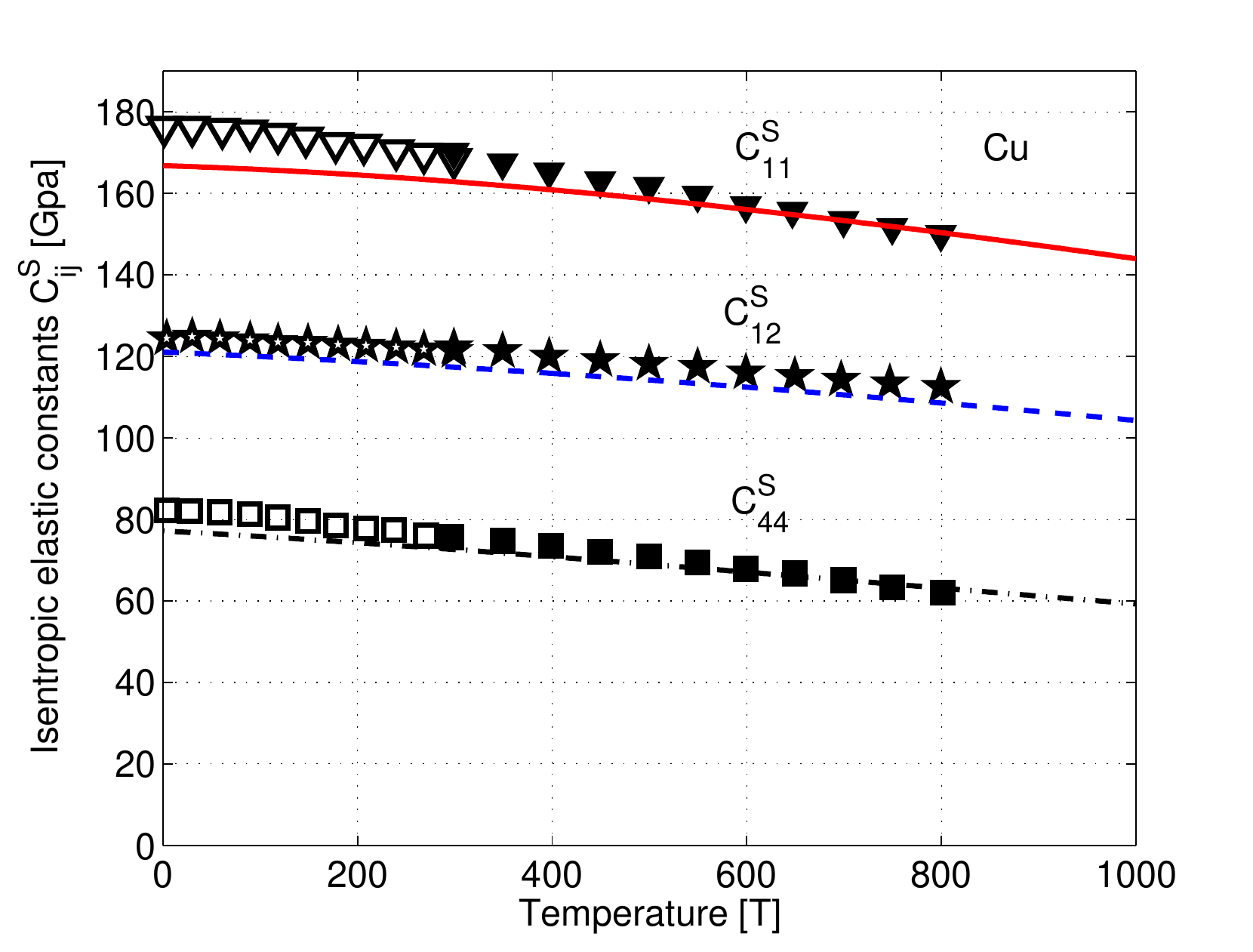}}
\caption{For single-crystal Cu, the present calculated isentropic elastic constants and corresponding experimental data. The solid, dashed, and dashed-dotted curves denote the present values of $C_{11}^{S}$, $C_{12}^{S}$ and $C_{44}^{S}$, respectively. The open symbols denote the corresponding values of ultrasonic measurements by Overton et al \cite{Overton}, and the filled symbols display those from the measurements of Chang and Himmel \cite{Chang}.}
\label{Cu}
\end{figure}

\begin{figure}
\scalebox{0.6}[0.6]{\includegraphics{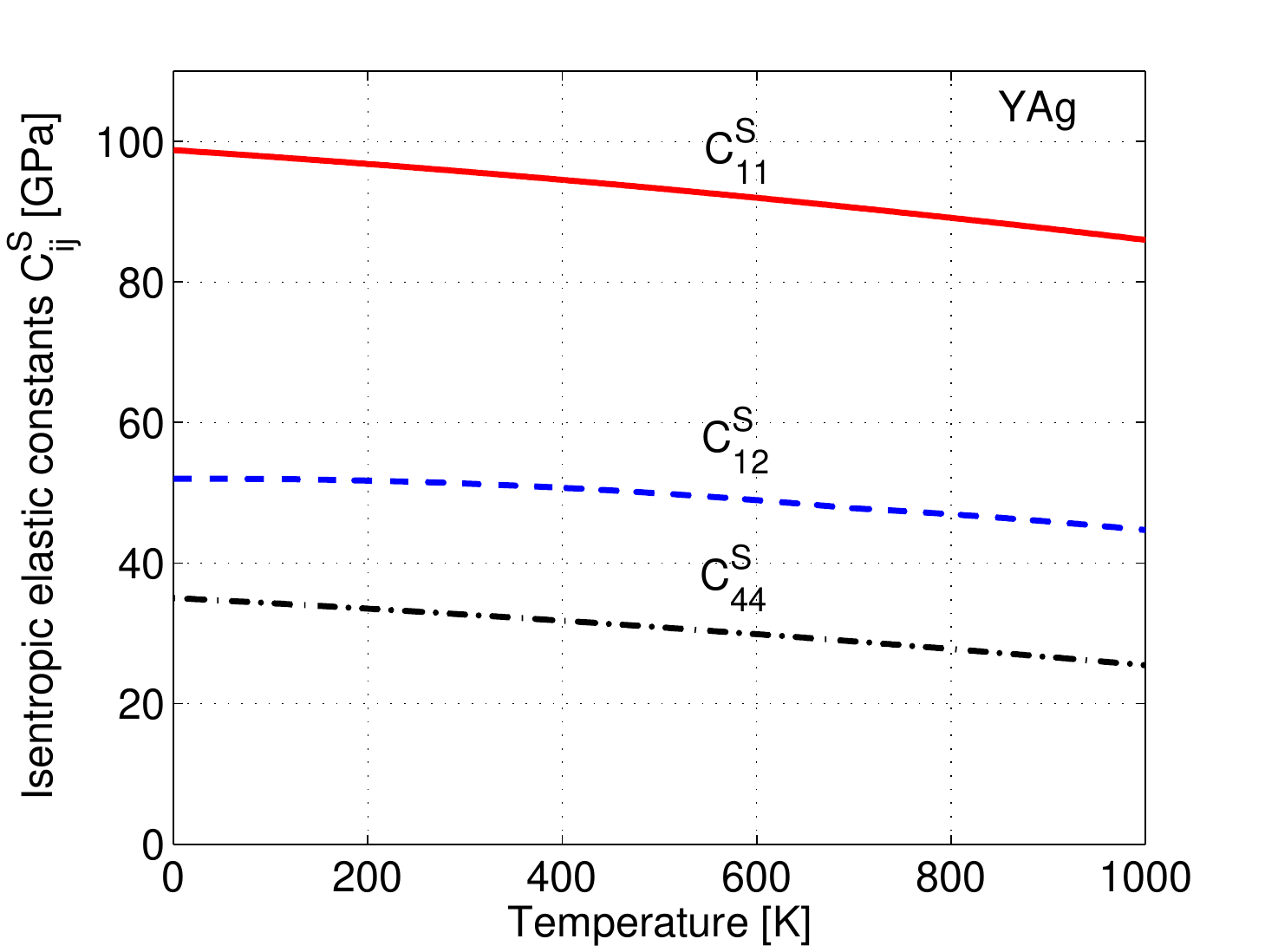}}
\caption{The predicted isentropic elastic constants as a function of temperature for YAg. The solid, dashed, and dashed-dotted curves denote the present values of $C_{11}^{S}$, $C_{12}^{S}$ and $C_{44}^{S}$, respectively. Note that $C_{11}>C_{12}$, $C_{11}>0$, and $C_{44}>0$ at all temperatures, which is consistent with stability of B2 lattice.}
\label{YAg}
\end{figure}

\begin{figure}
\scalebox{0.6}[0.6]{\includegraphics{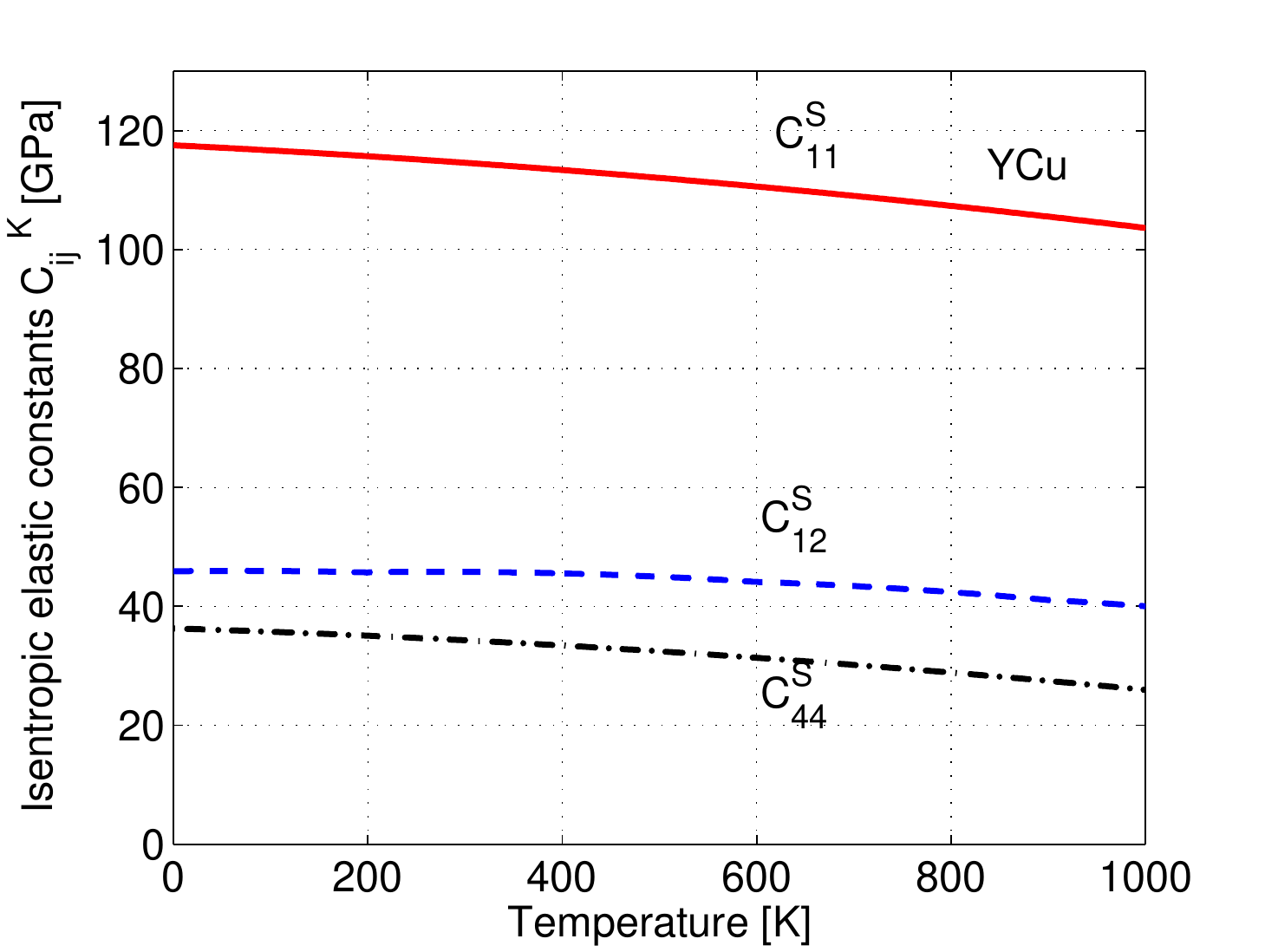}}
\caption{The predicted isentropic elastic constants as a function of temperature for YCu. The solid, dashed, and dashed-dotted curves denote the present values of $C_{11}^{S}$, $C_{12}^{S}$ and $C_{44}^{S}$, respectively. Note that $C_{11}>C_{12}$, $C_{11}>0$, and $C_{44}>0$ at all temperatures, which is consistent with stability of B2 lattice.}
\label{YCu}
\end{figure}

For the temperature dependence of the isentropic elastic moduli $C_{ij}^{S}$, we present in Figures \ref{Ag}-\ref{YCu} our calculated in the range 0-1000K at ambient pressure. Figures \ref{Ag} and \ref{Cu} show our findings for benchmark metals Ag and Cu respectively, accompanied
by available experimental data taken from ultrasonic measurements \cite{Neighbours,Overton,Chang}. Figures \ref{YAg} and \ref{YCu} give our prediction for the unknown values of temperature-dependent $C_{ij}^{S}$ for YAg and YCu, respectively.  Through the calculated values of isentropic elastic constants for Ag and Cu in comparison with the values measured by experiments,  we get a useful test of the accuracy of the method and the precision of our calculations of temperature dependence of elastic constants for YAg and YCu. The same DFT method had also been employed to calculate the elastic constants of Cu as a function of temperature by Wang et al \cite{Wang}, and our results show no discrepancy from their values. The overall observation is that all the calculated values of $C_{11}^{S}$, $C_{12}^{S}$, and $C_{44}^{S}$ decrease with increasing temperature, since thermal expansion may soften the elastic moduli at high T. At higher temperature, we also find that trend of $C_{ij}^{S}$ approach linearity. Considering our calculated results both for YAg and YCu, we find the values of $C_{11}^{S}$  decrease to the largest extant in the whole temperature range 0-1000K, and those of $C_{12}^{S}$ decrease least. $C_{11}^{S}$, $C_{12}^{S}$, and $C_{44}^{S}$ for YAg  decrease by 12.8GPa, 7.3GPa, and 9.6GPa, and those for YCu decrease by 13.9GPa, 5.9GPa, and 10.3GPa, respectively. The requirement of mechanical stability in a cubic crystal leads to the following restrictions on the elastic constants, $C_{11}-C_{12}>0$, $C_{11}>0$, and $C_{44}>0$ \cite{Nye}. Our results of the elastic constants as shown in Figure \ref{YAg} and \ref{YCu} satisfy these stability conditions in the range 0-1000K.

\begin{figure}
\scalebox{0.6}[0.6]{\includegraphics{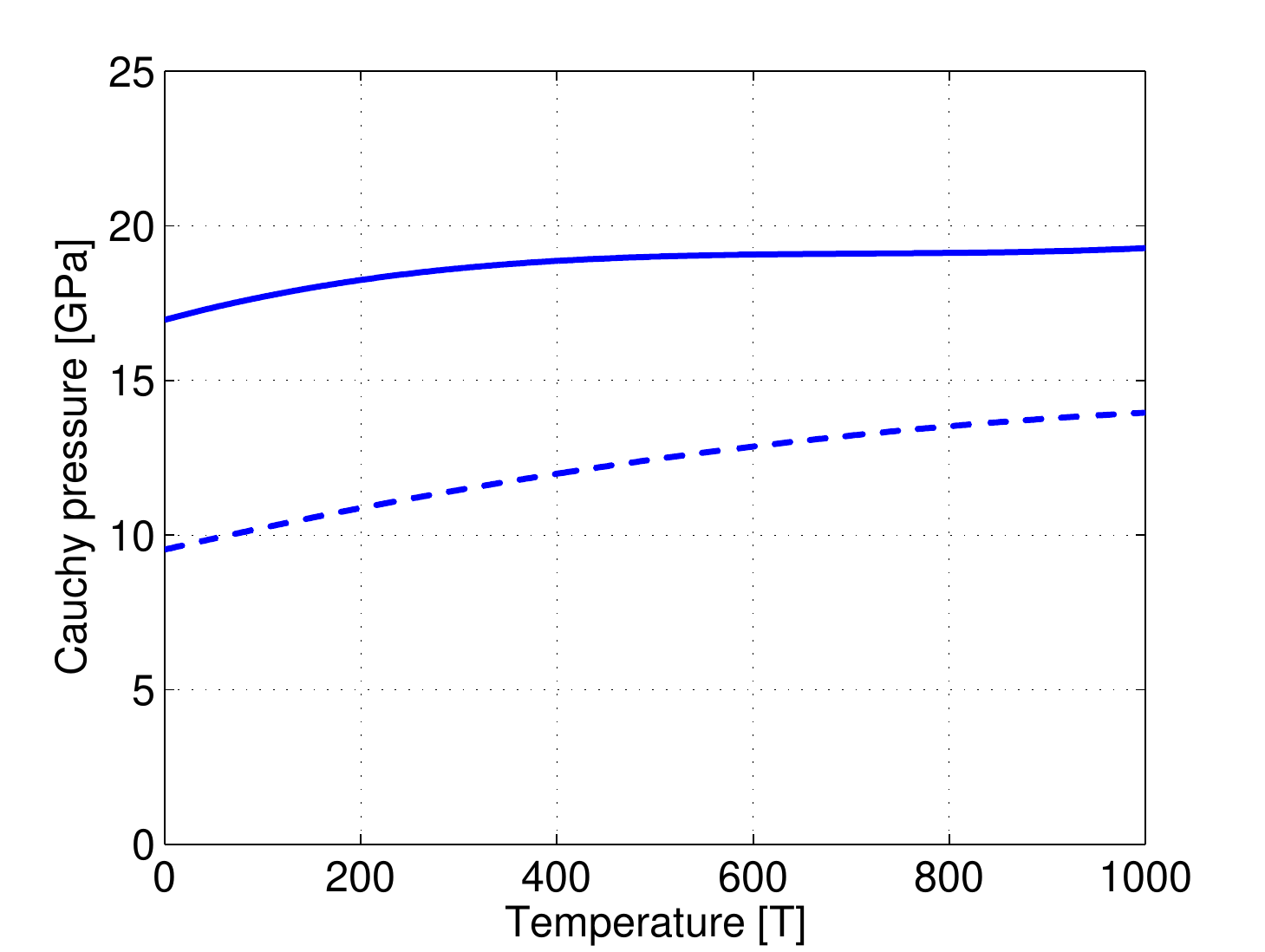}}
\caption{The predicted isentropic Cauchy Pressure defined by $P_{C}^{S}=C_{12}^S-C_{44}^S$ as a function of temperature. The solid and dashed curves denote the values for YAg and YCu, respectively.}
\label{cauchy}
\end{figure}

Next we discuss the temperature dependence of Cauchy pressure which was suggested that it could be used to describe the angular character of atomic bonding in metals and compounds by Pettifor \cite{Pettifor}. The isentropic Cauchy pressure $P_{C}^{S}$ is defined as
\begin{equation}
P_{C}^{S}=C_{12}^{S}-C_{44}^{S}.
\end{equation}
If the value of Cauchy pressure is more positive and bigger, the bonding of material is more metallic in character. In contrast, negative Cauchy pressure requires an directional character and low mobility in the bonding. Generally, for ductile materials such as metals Ag and Cu, the Cauchy pressures have positive values, while for brittle semiconductors such as Si, the Cauchy pressure is negative. The YAg and YCu are a new class of highly ordered and ductile intermetallics, especially for YAg in some cases exceeding 20\% ductility in tension \cite{Morris}. In table \ref{table},   the present results of Cauchy pressure  $T=0K$ and $T=300K$ have positive values for YAg and YCu, and agree well with the previous theoretical results at $T=0K$ and the experimental results at room temperature (about 300K). Figure \ref{cauchy} illustrates that the Cauchy pressure for YAg and YCu as a function of temperature. In the range of temperature 0-1000K, the Cauchy pressure of YAg is always greater than that of YCu, and those both for YAg and YCu increase with elevating temperature. Our calculated results demonstrate that high temperature may soften the directional character, and YAg and YCu will become more ductile while the temperature is increased.

\section{Conclusions}

In this work, we present the temperature-dependent elastic constants of two ductile rare-earth intermetallic compounds YAg and YCu with CsCl-type B2 structure by using a first-principles quasistatic approach, in which the static elastic constants as function of volume are determined by the first-principles DFT total-energy calculations within the framework of the method of homogeneous deformation and the volume-temperature relation determined by first-principles phonon calculations based on DFPT with quasiharmonic approach. To benchmark
the reliability results of the presented method, the comparison between
our calculated results for Ag and Cu with available experimental data has been performed and shows good agreements. In the whole temperature range 0-1000K, the elastic constants follow a normal behavior with temperature that those decrease with increasing temperature, and satisfy the stability conditions for B2 structures. The values of $C_{11}^{S}$ decrease to the largest extant, and those of $C_{12}^{S}$ decrease least. The  temperature dependence of Cauchy pressure is also discussed. The Cauchy pressures both for YAg and YCu increase with elevating temperature, and that of YAg is always higher than that of YCu. Our results mean that increasing  temperature may improve ductility of YAg and YCu.


 \vskip 2in

\footnotesize

\def\refname{{\large\bfseries References}}

\newpage

\begin{table}
\caption{The present calculated lattice constants $a$ ({\AA}), isentropic elastic constants $C_{ij}^{S}$ (GPa), and Cauchy pressure $P_{C}^{S}$(GPa) for YAg and YCu at $T=0K$ and $T=300K$ compared to previous computed results and experimental data. Note that all the previous calculated results are $T=0K$ values, while the experiments are room temperature values about $T=300K$.}
\begin{tabular}{cccc}
  \hline
   &Present calculation& Previous calculations & Experiment \\
   \hline
  YAg &   &   &    \\
  \hline
  $a$ & 3.645$^{a}$, 3.656$^{b}$  &3.634$^{c}$, 3.627$^{d}$ & 3.619$^{e}$\\
  $C_{11}^{S}$ &98.9$^{a}$, 95.7$^{b}$  &105.0$^{c}$, 102.5$^{d}$    &102.4$^{e}$ \\
  $C_{12}^{S}$ &52.4$^{a}$, 51.4$^{b}$  &50.0$^{c}$, 56.5$^{d}$    &54.0$^{e}$\\
  $C_{44}^{S}$ &35.5$^{a}$, 32.7$^{b}$  &37.0$^{c}$, 37.8$^{d}$   &37.2$^{e}$ \\
  $P_{C}^{S}$ &16.9$^{a}$, 18.7$^{b}$  &13.0$^{c}$, 18.7$^{d}$   &16.8$^{e}$ \\
  \hline
  YCu &   &   &   \\
  \hline
  $a$ &3.484$^{a}$, 3.492$^{b}$   &3.477$^{c}$, 3.472$^{d}$ &  3.477$^{e}$ \\
  $C_{11}^{S}$ &117.4$^{a}$, 114.6$^{b}$  &116.0$^{c}$, 113.6$^{d}$  &113.4$^{e}$ \\
  $C_{12}^{S}$ &45.9$^{a}$, 45.6$^{b}$  &47.0$^{c}$, 48.4$^{d}$ &48.4$^{e}$\\
  $C_{44}^{S}$ &36.3$^{a}$, 33.4$^{b}$  &35.0$^{c}$, 36.8$^{d}$ &32.3$^{e}$\\
  $P_{C}^{S}$ &9.6$^{a}$, 12.2$^{b}$  &12.0$^{c}$, 11.6$^{d}$   &15.9$^{e}$ \\
  \hline
\end{tabular}

\begin{tabular}{c}
  \leftline {${}^{a}$ This work at $T=0K$;} \\
  \leftline {${}^{b}$ This work at $T=300K$;} \\
  \leftline {${}^{c}$ Ref.\cite{Morris}, obtained from first-principles calculations at $T=0K$;} \\
  \leftline {${}^{d}$ Ref.\cite{Wu1}, obtained from first-principles calculations at $T=0K$;}\\
  \leftline {${}^{e}$ Ref.\cite{Morris}, obtained from experiments at $T=300K$.}\\
\end{tabular}
\label{table}
\end{table}

\end{document}